\documentstyle[aps,twocolumn,epsfig]{revtex}
\begin{document}
\title{Nuclear absorption of Charmoniums in pA and AA collisions}
\author{\bf A. K. Chaudhuri\cite{byline}}
\address{ Variable Energy Cyclotron Centre\\
1/AF,Bidhan Nagar, Calcutta - 700 064\\}
\maketitle
\begin{abstract}

We  have  analyzed the latest NA50 data on $J/\psi$ production in
pA and AA collisions. The $J/\psi$ production is assumed to be  a
two   step   process,   (i)   formation   of   $c\bar{c}$  pairs,
perturbatively calculable, and (ii) formation  of  $J/\psi$  from
the  pair,  a  non-perturbative  process,  which  is conveniently
parameterized. In a nuclear medium, as the $c\bar{c}$ pair passes
through the nuclear medium, it gain relative square momentum  and
some  of  the  pairs can gain enough square momentum to cross the
threshold for open charm meson, leading to suppression in nuclear
medium. Few parameters of the model were fixed  from  the  latest
high  statistics  NA50  pA  and  NA38  SU  total  $J/\psi$  cross
sectional  data.  The  model  then  reproduces   the   centrality
dependence of $J/\psi$ over Drell-Yan ration in 200 GeV/c S+U and
158 GeV/c Pb+Pb collisions.
\end{abstract}


\section{Introduction}

Recently,  in  Quark  Matter  2002,  NA50 collaboration presented
their analysis of the nuclear  absorption  of  $J/\psi$  in  high
statistics  450  GeV  pA collisions \cite{na50-a}. They estimated
the $J/\psi$ nucleon absorption  cross  section  ($\sigma^{J/\psi
N}_{abs}$) in the framework of Glauber model. High statistics 450
GeV  pA  data  yield  $\sigma^{J/\psi  N}_{abs} = 4.4 \pm 1.0$ mb
\cite{na50-a}.  They  also  estimate  a  common   $\sigma^{J/\psi
N}_{abs}$ from latest pA and NA38 200 GeV/c S+U data \cite{na38},
$\sigma^{J/\psi  N}_{abs}$  =4.4  $\pm$  0.5  mb.  The  extracted
absorption cross section is much smaller than the  earlier  value
of  6.4  $\pm$  0.8  mb  extracted  from fit to earlier NA50 data
\cite{na50-b} or 7.1 $\pm$ 3.0 mb obtained from a fit to NA38 S+U
data \cite{na38}.  Within  error,  the  S+U  cross  sections  are
compatible  with  pA  cross  sections. In Quark matter 2002, NA50
collaboration also presented their preliminary analysis  of  2000
Pb+Pb  run  \cite{na50-c}.  $E_T$ dependence of the $J/\psi$ over
Drell-Yan ratio, in the rapidity range of 2.9-4.5 was  presented.
Compared  to  1998  run  \cite{na50-d},  2000  data  are flatter,
suppression being more at low $E_T$ and less at high  $E_T$.  The
difference  may  be  attributed  to  the  difference in method of
analysis. In 1998, NA50 collaboration presented  combined  result
of  standard as well minimum bias analysis. In the 2000 run, only
the standard analysis was  performed.  Centrality  dependence  of
$J/\psi$  absorption is still anomalous in the sense that Glauber
model of nuclear absorption fails to explain the data.  The  1998
data  of  NA50  collaboration  \cite{na50-d}  were  analyzed in a
variety  of  models,  with  and   without   assumption   of   QGP
\cite{bl00,ch01,ca00,ch02,ch02a}.  We have shown that a QCD based
model, where $J/\psi$'s are  absorbed  in  nuclear  medium  could
explain the data \cite{ch02,ch02a,qiu98}. Parameters of the model
were  obtain  from a fit to then existing pp/pA/AA total $J/\psi$
cross section data \cite{na50-b}.

Preliminary  analysis  of  high  statistics pA data \cite{na50-a}
with the implication of less absorption of  $J/\psi$  in  nuclear
medium,  along  with  the changed suppression pattern observed in
the preliminary analysis of NA50 2000  Pb+Pb  run  \cite{na50-c},
necessitated   the   re-examination  of  the  QCD  based  nuclear
absorption model \cite{ch02}. In the  present  brief  report,  we
have  analyzed the latest NA50 data on pA/SU total $J/\psi$ cross
section \cite{na50-a} to obtain the parameters of the  QCD  based
nuclear  absorption model. The model was then used to explain the
centrality dependence of $J/\psi$ over  Drell-Yan  ratio  in  200
GeV/c   S+U   collisions   and  in  158  GeV/c  Pb+Pb  collisions
(preliminary data). As will be shown in the following,  once  the
model  parameters  are  fixed from $J/\psi$ total cross sectional
data, the QCD based nuclear absorption model  give  a  consistent
description  of the centrality dependence of $J/\psi$ suppression
in S+U collisions as well as in Pb+Pb collisions. The plan of the
paper is as follows: in section 2, we will briefly  describe  the
model  and  obtained  the  model  parameters  from  the  new high
statistics pA/AA NA50 data. In section 3, the NA50 data on  $E_T$
dependence  of  $J/\psi$ over Drell-Yan ratio in S+U and in Pb+Pb
collisions are analyzed.   Summary and conclusions are
drawn in section 4.

\section{$J/\psi$ production and suppression}

In  the  QCD  based  nuclear  absorption model \cite{ch02,qiu98},
$J/\psi$ production is assumed to be  a  two  step  process,  (a)
formation of a $c\bar{c}$ pair, which is accurately calculable in
QCD  and  (b)  formation  of a $J/\psi$ meson from the $c\bar{c}$
pair,  which  is  non-perturbative  but   can   be   conveniently
parameterized.  The $J/\psi$ cross section in $AB$ collisions, at
center of mass energy $\sqrt{s}$ is written as,

\begin{eqnarray} \label{1a}
\sigma^{J/\psi} (s) &&
=K \sum_{a,b} \int dq^2 \left( \frac{\hat \sigma_{ab \rightarrow
cc}} {Q^2} \right) \int dx_F \phi_{a/A}(x_a,Q^2) \\ \nonumber
&&   \phi_{b/B}(x_b,Q^2)   \frac{x_a   x_b}{x_a   +  x_b}  \times
F_{c\bar{c} \rightarrow J/\psi} (q^2), \end{eqnarray}

\noindent  where  $\sum_{a,b}$  runs over all parton flavors, and
$Q^2 = q^2 +4 m_c^2$. The  $K$  factor  takes  into  account  the
higher  order corrections. The incoming parton momentum fractions
are fixed by kinematics and are $x_a
=(\sqrt{x^2_F+4Q^2/s}+x_F)/2$               and              $x_b
=(\sqrt{x^2_F+4Q^2/s}-x_F)/2$.
$\hat \sigma_{ab \rightarrow c\bar{c}}$ are the sub process cross
section  and  are given in \cite{be94}. $F_{c \bar{c} \rightarrow
J/\psi}(q^2)$ is the transition  probability  that  a  $c\bar{c}$
pair  with  relative momentum square $q^2$ evolve into a physical
$J/\psi$ meson. It is parameterized as,

\begin{eqnarray} \label{4} F_{c \bar{c} \rightarrow J/\psi} (q^2)
= && N_{J/\psi} \theta(q^2) \theta({4m^\prime}^2 - 4 m_c^2 -q^2) \\
\nonumber   &&   (1   -   \frac{q^2}{{4m^\prime}^2   -   4  m_c^2
})^{\alpha_F}. \end{eqnarray}

\begin{figure}[h]                                  \vspace{-.5cm}
\centerline{\psfig{figure=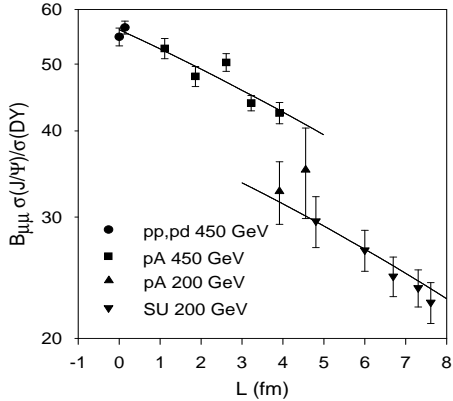,height=9cm,width=7cm}}
\vspace{-2.7cm} \caption{The experimental ratio of total $J/\psi$
cross section
 and  Drell-Yan  cross  sections in proton-proton, proton-nucleus
and nucleus-nucleus collisions. The fit to the data  obtained  in
the  QCD based nuclear absorption model is shown as solid lines.}
\end{figure}

In  a  nucleon-nucleus/nucleus-nucleus  collision,  the  produced
$c\bar{c}$ pairs interact with nuclear medium before  they  exit.
It  is  argued  \cite{qiu98} that the interaction of a $c\bar{c}$
pair  with  nuclear  environment  increases  the  square  of  the
relative  momentum between the $c\bar{c}$ pair. As a result, some
of the $c\bar{c}$ pairs can gain enough relative square  momentum
to   cross   the   threshold  to  become  an  open  charm  meson.
Consequently,  the  cross  section  for  $J/\psi$  production  is
reduced  in comparison with nucleon-nucleon cross section. If the
$J/\psi$ meson travel a distance $L$,  $q^2$  in  the  transition
probability  is  replaced  to $q^2 \rightarrow q^2 +\varepsilon^2
L$, $\varepsilon^2$ being the relative square momentum  gain  per
unit  length.  Parameters  of the model ($\alpha_F$,$KN_{J/\psi}$
and $\varepsilon^2$) can be fixed from experimental data on total
$J/\psi$ cross section in pA/AA  collisions.  Earlier  NA50  data
\cite{na50-b}   are   well  described  with  $KN_{J/\psi}$=0.458,
$\varepsilon^2=0.225 GeV^2/fm$ and $\alpha_F  =1.0$  \cite{ch02}.
As  mentioned in the beginning, NA50 collaboration presented high
statistics pA data on $J/\psi$ cross section. They have  measured
$B_{\mu\mu}  \sigma(J/\psi)/\sigma(DY)$.  In  Fig.1, experimental
data are shown as a function of  nuclear  length.  The  Drell-Yan
cross  sections donot have any A or alternately any L-dependence.
The observed $L$ dependence is then due to  $J/\psi$'s  only.  We
fit        the        data       with       two       parameters,
$N_{norm}=KN_{J/psi}/\sigma^{DY}_{NN}(nb)$  and  square  momentum
gain  factor  $\varepsilon^2$ ($\alpha_F$ being kept fixed at 1).
In Fig.1, the fit obtained to the data is shown. The two sets  of
data  at  200  GeV/c  and 450 GeV/c could be fitted with a common
square momentum gain factor, $\varepsilon^2$= 0.1875  $GeV^2/fm$,
a  value  20\% lower than the value obtained earlier \cite{ch02}.
Lowering  of  $\varepsilon^2$   indicate   less   absorption   of
$J/\psi$'s in nuclear medium, in agreement with the Glauber model
calculations.  While  the square momentum gain factor do not show
energy dependence, the evident energy  dependence  of  the  cross
section  ratios  shows  up  in  the  other parameter of the model
$N_{norm}$. We  obtain  $N_{norm}$  =  10.18  at  200  GeV/c  and
$N_{norm}$ = 4.43 at 450 GeV/c. The energy dependence of $J/\psi$
cross section being taken care of in the model (Eq.\ref{1a}), the
energy  dependence  of  $N_{norm}$  is due to the Drell-Yan cross
sections only. In the mass range, $2.9 > M >4.5$ GeV, the Craigie
parameterization \cite{craigie}, of the Drell-Yan cross  section,
$\sigma(DY)  \propto  e^{-14.9M/\sqrt{s}}$,  gives  for the ratio
$\sigma(DY)_{450GeV}/\sigma(DY)_{200GeV}$ = 2.1 -3.1,  consistent
with the presently obtained ratio of 2.29.

\section{$E_T$ dependence of $J/\psi$/DY ratio}

In  the  present  section,  we  analyzed  the $E_T$ dependence of
$J/\psi$ over Drell-Yan ratio in 200 GeV/c S+U collisions and  in
158  GeV/c  Pb+Pb  collisions.  Details of the calculation can be
found in ref.\cite{ch02}.

\begin{figure}[h]                                   \vspace{-0cm}
\centerline{\psfig{figure=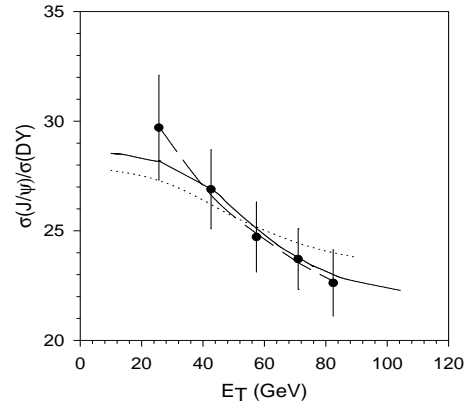,height=9cm,width=7cm}}
\vspace{-3cm}   \caption{The   transverse  energy  dependence  of
$J/\psi$ over Drell-Yan ratio in 200 GeV/c  S+U  collisions.  The
dashed and dotted lines are the fit obtained in the Glauber model
of  nuclear  absorption  with $\sigma^{J/\psi N}_{abs}$=7.1mb and
4.4 mb respectively. The solid line is the fit  obtained  in  the
QCD based nuclear absorption model.} \end{figure}

In       Fig.2,       NA38       experimental       ratio      of
$B_{\mu\mu}\sigma(J/\psi)/\sigma(DY)$   for   200    GeV/c    S+U
collisions   \cite{na38}  is  compared  with  the  present  model
calculation (solid line). The data  are  well  described  in  the
model.  For  comparison  purpose,  we  have  also  shown  the fit
obtained to the data in the Glauber model. The dashed line is the
Glauber model calculations with  $\sigma^{J/\psi  N}_{abs}$=  7.1
mb.  However,  quality  of  fit gets poorer if we use lower value
(4.4 mb) for $\sigma^{J/\psi N}_{abs}$ as extracted from the most
recent data (the dotted line). It is evident that in the  Glauber
model  of nuclear absorption, the transverse energy dependence of
$J/\psi$'s prefer larger absorption of $J/\psi$ in S+U collisions
than indicated in pA/AA total $J/\psi$ cross  section  data.  The
QCD  based  nuclear  absorption  model,  on  the other hand, give
consistent description of the data.

\begin{figure}[h]                                  \vspace{-.5cm}
\centerline{\psfig{figure=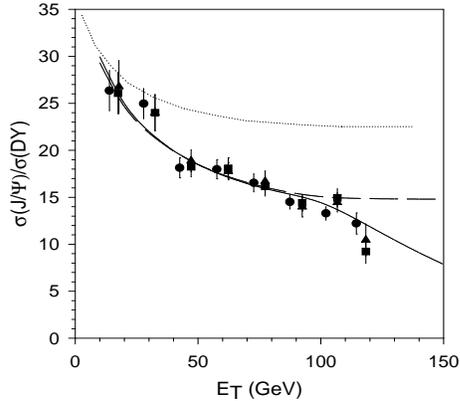,height=9cm,width=7cm}}
\vspace{-3cm}   \caption{NA50   preliminary   transverse   energy
dependence of $J/\psi$ over Drell-Yan ratio in  158  GeV/c  Pb+Pb
collisions. The dotted line is the Glauber model calculation. The
dashed line is the QCD based nuclear absorption model calculation
without  including  the  effect  of $E_T$ fluctuations. The solid
line  is  obtained   with   including   the   effect   of   $E_T$
fluctuations.} \end{figure}

Transverse      energy      dependence      of     the     ratio,
$B_{\mu\mu}\sigma(J/\psi)/\sigma(DY)$   in   158   GeV/c    Pb+Pb
collisions could not be explained in the Glauber model of nuclear
absorption.  The  preliminary  NA50 data \cite{na50-c}obtained in
2000, are flatter compared to 1996/1998 data \cite{na50-d}, still
the usual Glauber model of nuclear absorption fails  to  describe
the data. In Fig.3 the experimental data are shown along with the
Glauber  model  fit  to  it obtained with $\sigma^{J/\psi}_{abs}$
=4.4 mb. The Glauber model calculation (the  dotted  line)  agree
only  for the peripheral collisions. For more central collisions,
the model predict less absorption than in  experiment.  Also  the
Glauber  model  shows  saturation  at more central collisions, in
contrast to experiment. The 2nd drop in the ratio is  mainly  due
to  transverse  energy  fluctuations  at a fixed impact parameter
\cite{bl00,ch02b}. In the present model,  we  take  into  account
$E_T$  fluctuations  at  a  fixed  impact  parameter  $b$, by the
replacement: $L(b,s) \rightarrow L(b,s)E_T/<E_T>(b)$.  In  Fig.3,
the dashed is obtained in the QCD based nuclear absorption model,
without  incorporating  the $E_T$ fluctuation effect. It explains
the data up to 100 GeV,  the  knee  of  the  $E_T$  distribution,
beyond  which  only,  $E_T$  fluctuations are important. If $E_T$
fluctuation is included, the data are  explained  throughout  the
$E_T$  range  (the  solid  line). Consistent description of $E_T$
dependence of $J/\psi$ over Drell-Yan ratio in 200 GeV/c S+U  and
158 GeV/c Pb+Pb collisions clearly demonstrate that the QCD based
nuclear   absorption   model,  together  with  transverse  energy
fluctuation could explain consistently  the  $J/\psi$  absorption
data at SPS energy.

\section{summary and Conclusions}

To summarize, we have analyzed the latest NA50 data \cite{na50-c}
on $J/\psi$ suppression obtained in 2000 Pb+Pb run. The data were
analyzed  in the QCD based nuclear absorption model. The $J/\psi$
production is assumed to be two step process, (i)  production  of
$c\bar{c}$  pair, purturbatively calculable and (ii) formation of
$J/\psi$ from the $c\bar{c}$ pair. In a nuclear  medium,  due  to
random  collisions  with  medium, relative square momentum of the
$c\bar{c}$ increases and some of the pair may gain enough  square
momentum  to  cross the threshold for open charm meson. We obtain
the parameter, square  momentum  gain  factor  per  unit  length,
$\varepsilon^2$,   from   a   fit   to   the   latest  NA50  data
\cite{na50-a}, and obtain $\varepsilon^2$=0.1875 $GeV^2/fm$.  The
value  is lower than the value $\varepsilon^2$= 0.225 $GeV^2/fm$,
obtained in our earlier work \cite{ch02}. There,  we  had  fitted
the  earlier version of the NA50 data \cite{na50-b}. Reduction in
$\varepsilon^2$ indicate that the $J/\psi$'s are less absorbed in
nuclear medium. This is in accordance with  the  $J/\psi$-nucleon
absorption    cross    section,    extracted    from   the   data
\cite{na50-a,na50-b}. $\sigma^{J/\psi  N}_{abs}$  extracted  from
the  earlier  version  of  the NA50 data \cite{na50-b} was large,
$\sigma^{J/\psi N}_{abs} = 6.4 \pm 0.8$ mb, compared  to  the  value
extracted from the latest data set \cite{na50-a}, $\sigma^{J/\psi
N}_{abs} = 4.4 \pm 1.0$ mb. With the new model parameter, the model
could   explain   the  centrality  dependence  of  $J/\psi$  over
Drell-Yan ratio in 200 GeV/c S+U collisions as well as the latest
NA50 data on the centrality dependence of $J/\psi$ over Drell-Yan
ratio in 158 GeV/c Pb+Pb collisions.

\end{document}